\documentclass[manuscript]{aastex}

\shorttitle{Kelvin-Helmholtz Instability in the Solar Corona}
\shortauthors{Ofman and Thompson}

\begin{document}


\title{SDO/AIA Observation of Kelvin-Helmholtz Instability in the Solar Corona}


\author{L. Ofman\altaffilmark{1,2,3} and B. J. Thompson\altaffilmark{2}}


\altaffiltext{1}{Catholic University of America, Washington, DC
20064} \altaffiltext{2}{NASA Goddard Space Flight Center, Code 671,
Greenbelt, MD 20771} \altaffiltext{2}{Visiting, Department of
Geophysics and Planetary Sciences, Tel Aviv University, Tel Aviv,
Israel}


\begin{abstract}
 We present observations of the formation, propagation and decay of
vortex- shaped features in coronal images from the Solar Dynamics
Observatory (SDO) associated with an eruption starting at about
2:30UT on Apr 8, 2010. The series of vortices formed along the
interface between an erupting (dimming) region and the surrounding
corona. They ranged in size from several to ten arcseconds, and
traveled along the interface at 6-14 km s$^{-1}$. The features were
clearly visible in six out of the seven different EUV wavebands of
the Atmospheric Imaging Assembly (AIA). Based on the structure,
formation, propagation and decay of these features, we identified
the event as the first observation of the Kelvin-Helmholtz
(KH) instability in the corona in EUV. The interpretation is supported by linear analysis and by MHD model of KH instability. We conclude that the
instability is driven by the velocity shear between the erupting and
closed magnetic field of the Coronal Mass Ejection (CME). The shear flow driven instability can play an important role in energy transfer processes in coronal plasma.
\end{abstract}



\section{Introduction}

The Kelvin-Helmholtz (KH) instability, produced by two fluids
undergoing differential shearing motion across an interface, was
described by \citet{Kel1871} and \citet{Hel1868} almost
one and a half centuries ago. The KH instability is observed over a
wide range of gaseous, fluid, and  plasma regimes on the Earth
 and in space. Chains of vortex-shaped features have been observed in
clouds on the Earth, and in the atmospheric cloud belts of Jupiter
and Saturn that exhibit shearing wind motions. These vortices were
interpreted as signatures of the KH instability. KH vortices have
been observed in the Earth's aurora \citep{Far94} and the
magnetosphere, where there is evidence that the instability plays an important 
role in the energy transport of solar wind into the magnetosphere
\citep{Has04}. Observations have also been made in the magnetospheres
of Mercury \citep{Sla10}, Saturn \citep{Mas10}, and Ganymede
\citep{Kiv98}. Recently, evidence of KH development in solar
prominence material was found in observations from the Hinode Solar
Optical Telescope \citep{Ryu10,Ber10}.

The growth rate of the instability was calculated analytically
for sheared flow in an idealized fluid  \citep{Cha61}. This instability can also
occur in magnetized plasma of the solar corona; in this case the
magnetic field component along the direction of the shear can have a
stabilizing effect \citep{Cha61} and can also affect the growth rate
and structure of the vortices \citep{LS96}.Non-uniform (sheared) magnetic field across the velocity shear layer  can have stabilizing or
destabilizing effect on the KH mode, depending on the properties of
the magnetic shear \citep[e.g.,][]{OCM91}. The KH instability has been studied
theoretically in the context of coronal plasmas, where it is
believed to play an important role in the transition to turbulence
and plasma heating \citep{HP83,Ofm94,Kar94}.KH instability was also proposed as a mechanism that produces the observed quasi-periodic pulsation in flare high energy emission \citep{OS06}. In this study we report the
first observation of KH instability in the solar corona in EUV
coronal lines by SDO/AIA. The interpretation of observations is supported by linear analysis and MHD model of KH instability. The paper is organized as follows: \S~2 is devoted to the description of SDO/AIA observations, \S~3 described briefly the theory of KH instability, \S~4 is devoted to numerical result of MHD model,  and the conclusion are in \S~5.

\section{Observations}

The Atmospheric Imaging Assembly (AIA) on SDO is an array of four
telescopes that captures images of the Sun's atmosphere out to
1.3$R_\odot$ in ten separate wave bands. Seven of the ten channels
are centered on EUV wavelengths: 94\AA\ (Fe XVIII), 131\AA\ (Fe
VIII, XX, XXIII), 171\AA\ (Fe IX), 193\AA\ (FeXII, XXIV), 211\AA\
(Fe XIV), 304\AA\ (He II) and 335\AA\  (Fe XVI), representing a
range of effective temperatures from the upper chromosphere to the
corona. The images are 4096$\times$4096 square with a pixel width of
0.6\arcsec.

\begin{figure}
\includegraphics[scale=.78]{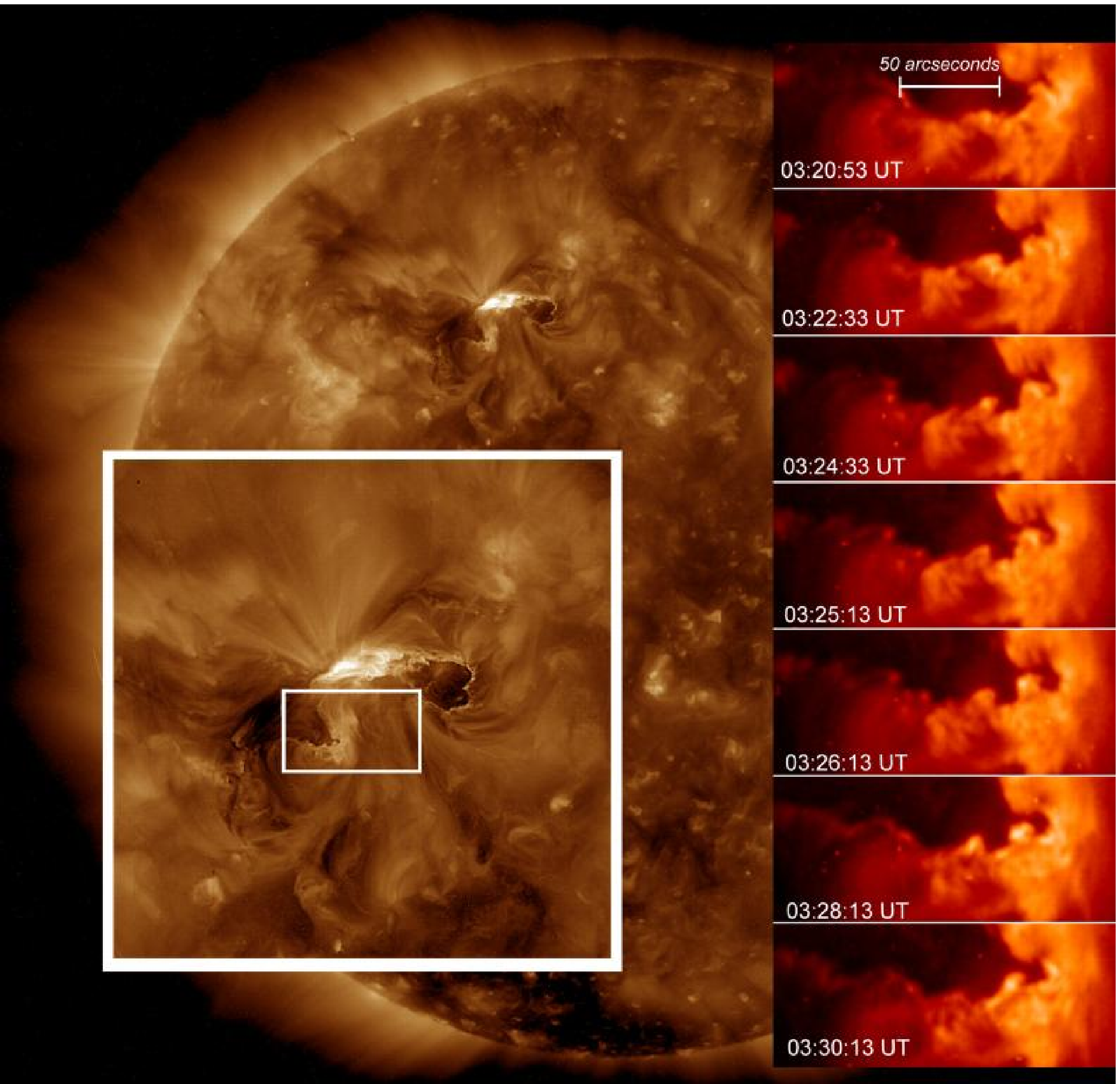}
\caption{The site of the CME eruption that begins around 02:34 UT on
8 April 2010 is shown in a full-disk 193\AA\ image.  The large white
box is a zoomed-in view of the erupting structure, which exhibits
dark regions of evacuated material as well as the flaring active
region. The smaller box highlights the boundary between the dark
region and the surrounding corona, which is where the KH vortices
were observed.  The sequence of frames on the right shows the
temporal evolution of the vortices in 211\AA\ images (see the
accompanying animations of this event). We found that the KH
features had the clearest visibility in the 211\AA\ images.}
\label{context:fig}
\end{figure}

SDO was launched on 11 February 2010, and on 8 April 2010 the
observatory was still undergoing its post-launch commissioning
phase. AIA was taking a full set of images every 20 seconds in
preparation for the transition to nominal science operations which
would have a 10- to 12-second cadence. On 8 April 2010 beginning at
02:34 UT, AIA observed a flare and coronal mass ejection coming from
an active region located 16$^{\circ}$ E and 29$^{\circ}$ N of
heliographic disk center. The flare peaked at 03:25 UT at a flux of
B3.7 on the GOES magnitude scale, and STEREO SECCHI measured the
CME's speed to be around 500 km s$^{-1}$. In addition to the flare,
the primary indicator of the CME in the AIA images was the formation
of large-scale dimming regions adjacent to the active region (see
Figure~\ref{context:fig}). These dimming regions are the site of
plasma evacuated by the eruption.

Along the boundary of one of the dimming regions, we identify the formation, propagation and decay of vortex-shaped
features ranging from several to ten arcseconds in size. They
traveled at speeds ranging from $\sim6 - 14$ km s$^{-1}$ along the
boundary, which we identify as the shearing interface between
erupting magnetic fields and the surrounding, non-erupting corona.
Based on the structure, evolution and decay of these features and supported by theoretical results (see below), we
believe that these are the first observations of the Kelvin-
Helmholtz instability in the corona in EUV.

The vortex-shaped features are clearly visible in six of AIA's EUV
wavebands, and are partially visible in the 94\AA\ images. The
corona imaged in the 94\AA\ is the hottest of the AIA wavebands
(around 6 MK), while the plasma at the dimming boundary is closer to
typical coronal temperatures of 1-2MK. The observed vortices are
small ($\sim$ 7000 km) and evolve rapidly, disappearing on
timescales of tens of minutes. In Figure~\ref{context:fig} we show
the solar disk observed in the EUV 171\AA\ waveband. The outer box
shows the location of the CME eruption, while the inner box shows
the boundary of the dimming region where the vortices were observed.
In the right panel we highlight the evolution of the vortices in a
sampling of 211\AA\ images ranging from t = 3:20:53 UT to t =
3:37:53 UT (see the accompanying animations of this event). The
vortices propagate from right to left at speeds ranged from $\sim6 -
14$ km s$^{-1}$. All in all, aspects of these vortices were visible
for more than an hour.

\begin{figure}
\includegraphics[scale=.60]{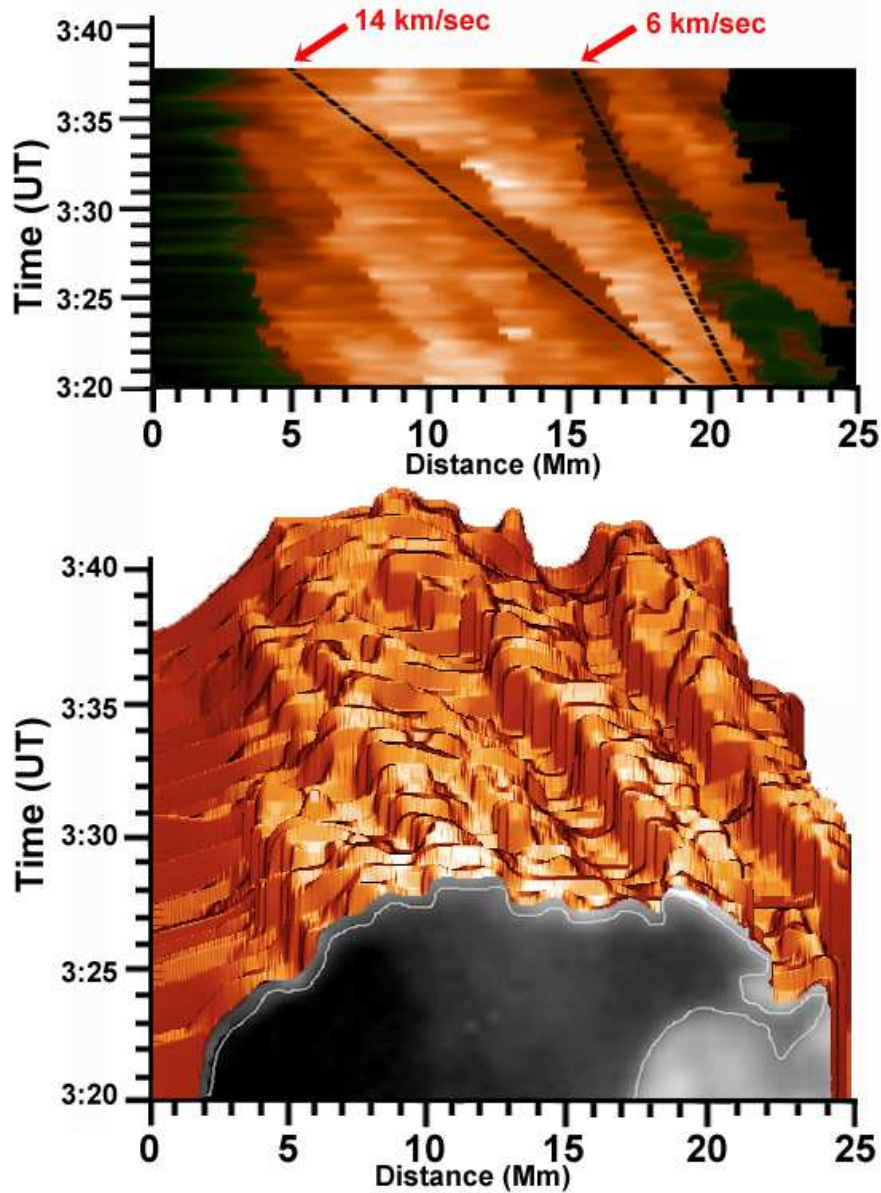}
\caption{The lower image shows the motion along the interface
between the evacuated (dark) corona and the surrounding non-eruptive
region. The interface was outlined in 51 separate 211\AA\ images,
and the results are shown in contour plot of space vs. time. (Note
that in this plot, the image coordinates are flipped vertically
relative to the actual observations.) The top image is the
projection of the lower plot onto the x-y plane.  The slopes that
correspond to motions of 6 km s$^{-1}$ and 14 km  s$^{-1}$ are indicated with
black dashed lines.} \label{contour:fig}
\end{figure}

The first sign of vortex formation is at 3:00:13UT, and the flow of
the vortices along the boundary was fully developed by 3:13:13UT. The
vortex and associated features lasted until about 4:47:13UT. In order
to study the motion of the vortices in Figure~\ref{contour:fig} we
show the interface boundary stacked in time, where the $x$-axis is
the location across the solar coordinate in arcseconds, and the
$y$-axis is time in seconds (lower panel). Cuts were taken from a
total of 51 images in 211\AA\ from a period of 3:20UT to 3:40UT,
which corresponds to the period in which the vortex motion was most
visible. The average difference between consecutive images was 20-25
seconds. In order to obtain the speed relative to the curved
interface the stacks are shown along the coordinate of the interface
$s$ on the $x$-axis, and time on the $y$-axis. The dashed lines in
the figure indicate slopes corresponding to speeds of 6 km s$^{-1}$
and 14 km s$^{-1}$, after compensating for the velocity of solar
rotation.

The motion of the magnetic loops associated with the erupting CME on
April 8, 2010 are shown in Figure 3. The five frames show difference
images that were created from consecutive pairs of 211\AA\ images
taken at $t=$2:51:13, 2:55:13, 3:00:33, 3:08:33, and 3:13:55UT. The
AIA image of the region at 3:13:53UT is shown for context. The
difference images enhance the structure and the evolution of the
loops. We have estimated the upper limit of the speed of the shearing field motion in
the initial stages of the eruption by tracking several loop features
as a function of time, and found that their velocity was $\sim$ 20
km s$^{-1}$. We assume that this velocity is close to the velocity
at which the material is evacuated from the dark regions, and that the projection effects are small. Thus, we
can estimate the relative velocity between the non-erupting
(stationary), and erupting region. By observing the width of the
interface between the two regions before the formation of the
vortices we estimate the thickness of the interface as 1-3 pixels.
Thus, on average the interface is $\sim$800 km thick, and the
velocity must change from zero to $\sim20$ km s$^{-1}$ over this
thickness. This is the initial driving shear of the KH instability
(see below).

The first indication of the appearance of the dimming structure is
around 2:42:53UT, and it's fairly well defined by 2:51:13UT. The
dimming structure expands and reaches the location where the KH
vortices eventually develop. The first image where we see any motion
along the KH front is 3:13:53UT. Thus, we can roughly estimate that
the growth time of the KH instability to the nonlinear stage is on the order of 14
minutes.

\section{Kelvin-Helmholtz instability}

The KH instability in fluids or plasmas arise when there is a velocity
discontinuity or finite thickness velocity shear over an interface between two regions \citep{Kel1871,Hel1868,Cha61} , and it is observed in many natural phenomena as well as in laboratory fluids or plsamas. In the initial linear stage the
KH instability exhibits exponential growth, followed by nonlinear
saturation and the formation of the typical KH vortices on the
interface between the fluids, thus broadening the interface and
reducing the magnitude of the shear. The kinetic energy of the shear is converted to the kinetic energy of the vortices that further transport it to smaller scales. In magnetized plasma the
orientation of the magnetic field has an important role on the
growth of the KH instability \citep{Cha61}. When the magnetic field
has significant component parallel to the interface, the growth of
the KH instability can be suppressed. The strength of the magnetic
field affects the size and the complexity of the vortices produced
by KH instability in magnetized plasma. The linear growth rate of KH
instability in an incompressible fluid with a discontinuous velocity
jump between the two regions is \citep{Cha61,Fra96}
\begin{eqnarray}
&&\gamma=\frac{1}{2}|\mbox{\bf k}\cdot\mbox{\bf $\Delta$ V}|[1-(2V_A
\mbox{\bf \^{k}}\cdot \mbox{\bf \^{B}})^2/(\mbox{\bf
\^{k}}\cdot\mbox{\bf $\Delta$ V})^2]^{1/2},
\label{kh_growth:eq}
\end{eqnarray} where $\mbox{\bf $\Delta$ V}$ is the velocity jump
over the interface, $\mbox{\bf B}$ is the magnetic field, $k$ is the
wavevector, $V_A=B/\sqrt{4\pi\rho}$ is the Alfv\'{e}n speed, and
$\rho$ is the density. When the interface region is of a finite
width the growth rate is reduced compared to the case with
discontinuous jump in velocity.

Based on the parameters of the observed velocity shear in the range 6-14 km
s$^{-1}$, wavelength of 7000 km (based on the size of the initial
ripples), and assuming that the velocity jump is discontinuous since
it is much smaller than the wavelength, we can estimate the upper limit of the
linear growth rate of the KH instability using the above equation: $\gamma_{KH}\approx
0.003-0.006$ s$^{-1}$. For simplicity, we have assumed that the magnetic
field has no parallel component to the interface in Equation~\ref{kh_growth:eq}. The nonlinear
saturation is reached at time scale of several $\gamma^{-1}$ or in
about 10 minutes, is in agreement with the observed time scale of
the evolution. We relax some of the simplifying assumptions in the MHD model. Note that the KH instability is not strongly
suppressed, implying that the magnetic field was mostly radial
(i.e., perpendicular to the plane of the velocity shear) at the
interface. This is in agreement with the observations of the
erupting loops, which stretch the magnetic field and evolve to a
nearly radial orientation as the eruption proceeds.

\section{Numerical Model}

Here, we describe the results of numerical 2.5D (i.e., two spatial dimensions and 3 components of magnetic field and velocity) MHD model of KH instability as a simplified model the magnetic and velocity structures seen by AIA. Numerical studies have been developed in the past to model the behavior of KH instability in a magnetized plasma, with initial velocity shear and weak magnetic field in the plane of the shear using a 2.5D MHD model \citep[e.g.,][and similar studies]{MP82,Miu84,Fra96}. Here, we solve the compressible resistive MHD equations in the 2D Cartesian $x-y$ plane, without gravity (since it is perpendicular to the plane), and isothermal energy equation \citep[see,][for the description of the equations and the normalization]{OT02} and with the following initial state (Figure~\ref{KH_init:fig}):
\begin{eqnarray}
&&V_{x0}(y)=(V_0/2){\rm tanh}(y/a)\\
&&\rho_0(y)=\frac{1}{2}(\rho_{max}+\rho_{min})-\frac{1}{2}(\rho_{max}-\rho_{min}){\rm tanh}(y/a),
\end{eqnarray} where $a$ is the width of the shear layer in $V_{x0}$ and $\rho_0$, and the density is varied similarly between $\rho_{max}$ and $\rho_{min}$, and $a=800$ km, guided by AIA observations. In order to determine the density ratio between the bright and dark (evacuated) regions across the interface we have examined the emission ratio in EUV AIA images for several wavebands. We found that the emission ratio is on average $\sim$5 with a range of 3-10. Since the density is proportional to the emission square for the same column depth in the line of sight, we have used $\rho_{max}/\rho_{min}=5^{1/2}$ in the model to represent the average observed value. We have also performed runs with density ratios of 2 and 3 and found similar results, with the growth rate of KH instability decreasing by $\sim$15\% with the density ratio in this range. The initial magnetic field is taken to be $\mbox{\bf B}_0=B_{x0}\hat{x}+B_{z0}(y)\hat{z}$, with constant $B_{x0}$, and the form of $B_{z0}(y)$ determined from pressure balance over the interface, i.e., in normalized units $B_{z0}(y)^2/2+\beta\rho_0(y)=const.$ Since we assume that the magnetic field is opened by the erupting CME we take the value of $B_{x0}\approx 0.04 <B_{z0}>$ resulting in small Alfv\'{e}n speed in the $x-y$ plane, $V_{A,xy}=B_{x0}/\sqrt{4\pi\rho}$ compared to the total Alfv\'{e}n speed $V_{A}=B/\sqrt{4\pi\rho}$, and the value of $V_0=5V_{A,xy}$. The value of the magnetic field in the observations is unknown. However, the results of the model are not sensitive to the magnitude of $B_{z0}$, and $B_{x0}$ as long as $V_{A,xy}<V_0$. The boundary conditions are periodic in the $x$-direction, and open in the $y$-directions. The size of the modeled region in the $y$-direction is $10a$, and in the $y$-direction $\sim 5\pi a$ corresponding to the wavelength of the fastest growing mode \citep[e.g.,][]{MP82}. The time is in units $\tau_A=a/V_{A,xy}$. The calculations are  initialized with a small amplitude fundamental harmonic perturbation in velocity in the $x-y$ plane. The MHD equations are solved using the modified Lax-Wendroff method with 4th order artificial viscosity term on $260\times260$ grid. The resistivity does not play a significant role on the time scale of the evolution of KH instability that grows in few Alfv\'{e}n times, and the Lundquist number is set to $10^4$.

The results of the 2.5D MHD model of KH instability are shown in Figure~\ref{rh_v:fig} at $t=17\tau_A$ in the nonlinear stage. The density structure and the corresponding velocity fundamental vortex generated by the KH instability are evident.  The structure and the temporal evolution of the density corresponds to the traveling structures seen in the time sequence of the SDO/AIA EUV emission in Figure~\ref{context:fig} and in the corresponding animations of the data and the model. The model shows that later evolution leads to formation of smaller scale vortices, and the development of low density 'bubbles' in the high density region, similar to observational features (see the animations). Although, the present model is for simplified 2.5D configuration, it reproduces the main observational features, supporting the interpretation of the observations in terms of KH instability.

\section{Discussion and Conclusions}

We report the first EUV observations of the KH instability in the
corona, which occurred along the interface between erupting and
non-erupting material associated with the CME of April 8, 2010. The
KH instability features are small (7000 km) and evolve on timescales
of seconds, so it would be difficult to observe this phenomenon with
an imager that has a lower resolution and cadence than SDO/AIA. The
detailed formation and evolution of KH vortices is observed and
analyzed. It was found that features along the velocity shear first
exhibit the linear growth stage, but the instability quickly reaches
nonlinear saturation (within tens of minutes), exhibiting the
typical formation of KH vortices. 

The growth and the evolution of
the instability is compared to theoretical predictions in the linear
stage, and to the results of 2.5D MHD model in the nonlinear stage, and good qualitative
agreement is found. In particular, we find that the modeled vortices and associated density structures reach the saturated, nonlinear stage, and the dynamics of the features is qualitatively similar to the observed evolution in AIA images.

The detection of the KH instability in the solar corona further shows that KH instability is a ubiquitous process in natural fluids and plasmas that can
play an important role in energy transport in the corona on all
scales. Theoretical studies find that KH instability is 
important in dissipation of free energy in shear flows and jets, and in the transition to turbulence. KH instability can also  occur
on small scales in solar coronal plasma in the presence of shear 
flows, leading to enhanced dissipation of waves and super-Alfv\'{e}nic jets produced
by impulsive events (such as flares) in the solar corona, and facilitate the heating of the solar
coronal plasma.

\acknowledgments We are grateful to SDO/AIA team for providing the
data used in this study. LO was supported by NASA grants NNX08AV88G,
NNX09AG10G.

\clearpage



\begin{figure}
\includegraphics[scale=.80]{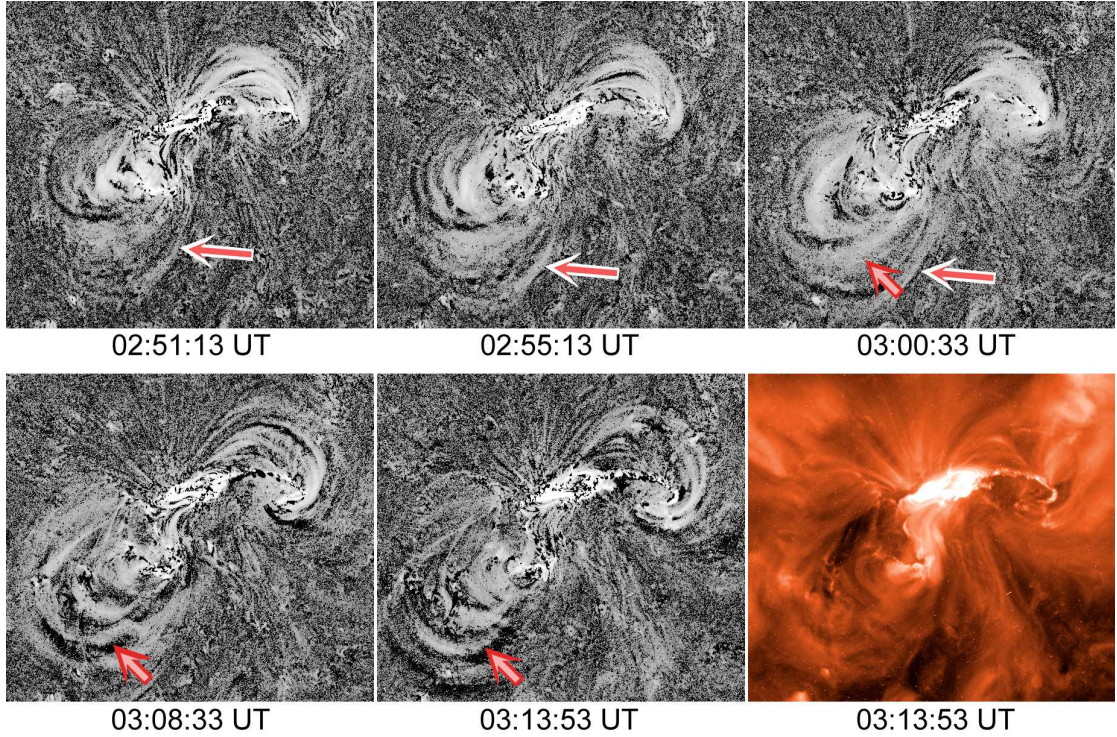}
\caption{The first five panels are difference images that were
created by subtracting consecutive pairs of 211\AA\ images.  The
different types of arrow indicate the motions of two separate loops
associated with the erupting CME shown in Figure ~\ref{context:fig}.
The lower right panel is a 211\AA\ AIA image of the eruption region
at 03:13:5UT, shown for context.  The displacement motion of several
different loops indicated that the motion near the arrows was around
20 km s$^{-1}$. } \label{loop_motion:fig}
\end{figure}

\begin{figure}
\centerline{\includegraphics[scale=1.2]{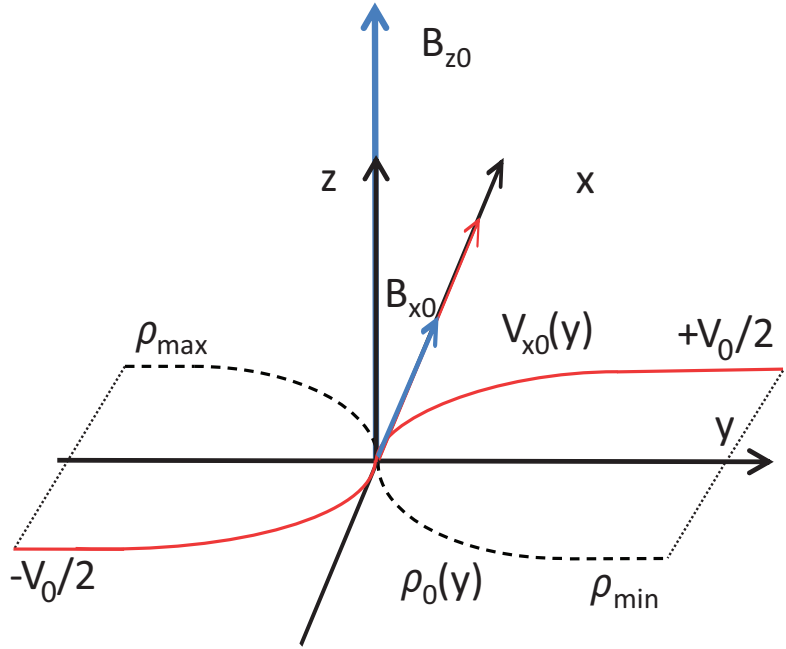}}
\caption{A sketch of the initial state used in 2.5D MHD model of the KH instability. The $x$-direction is parallel to the interface. The blue arrows show the magnetic field components (not to scale), where $B_{z0}$ represents the strong radial magnetic field components associated with the erupting CME, and $B_{x0}$ is the much weaker transverse field component. The red arrow shows the direction of the initial flow, the curves show schematically the variation of $V_{x0}(y)$ (red), and density $\rho_0(y)$ (dashes) across the interface.}
\label{KH_init:fig}
\end{figure}

\begin{figure}
\centerline{\includegraphics[scale=0.9]{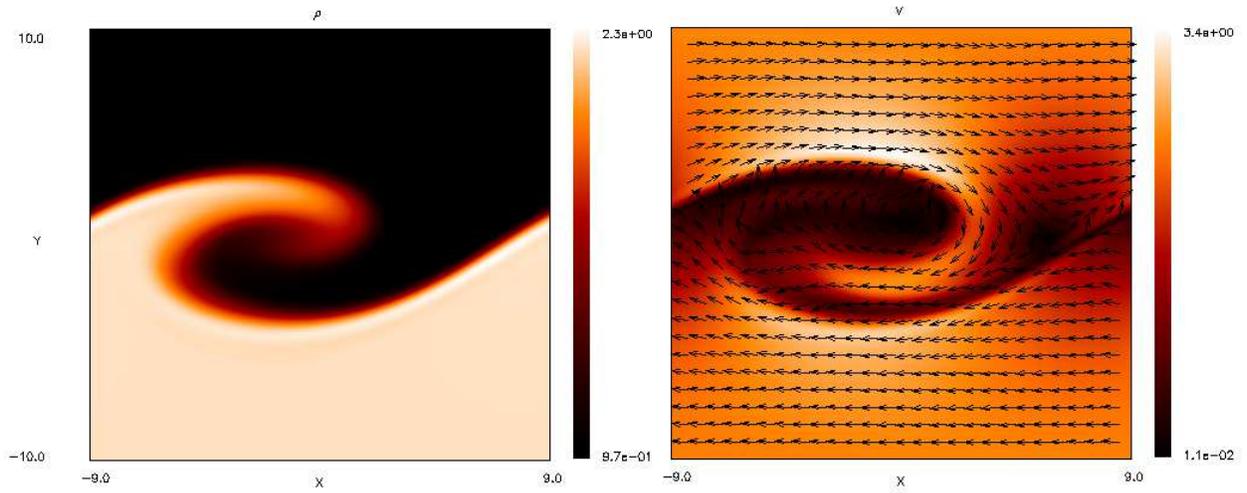}}
\caption{The density (left panel) and velocity (right panel) structures of the fully developed nonlinear KH instability at $t=17\tau_A$ obtained with the 2.5D MHD model. The density is in units of $\rho_{min}$. The arrows show the direction of the flow, and the magnitude is in units of $V_{A,xy}$ (see text). The animation of the density and velocity evolution is enclosed in the electronic version of this paper.}
\label{rh_v:fig}
\end{figure}

\end{document}